\documentclass[useAMS,usenatbib]{mn2e}
\usepackage[utf8]{inputenc}
\usepackage {graphicx,color}
\usepackage{xcolor}
\usepackage[english]{babel}
\usepackage{graphicx,natbib,multirow,topcapt}
\usepackage{amsmath,amssymb, mathtools}

\def\msun{M_{\odot}}
\def\mj{M_{{\rm J}}}
\def\mi{M_{{\rm i}}}
\def\ti{t_{{\rm i}}}
\def\zi{z_{{\rm i}}}
\def\ri{R_{{\rm i}}}
\def\di{\delta_{{\rm i}}}

\def\zf{z_{{\rm f}}}
\def\tf{t_{{\rm f}}}

\def\mf{M_{{\rm f}}}

\def\zrei{z_{{\rm rei}}}
\def\zprei{z_{{\rm prei}}}

\def\tita{t_{{\rm ita}}}
\def\zita{z_{{\rm ita}}}
\def\rita{r_{{\rm ita}}}

\def\rta{r_{{\rm ta}}}
\def\rin{r_{{\rm in}}}
\def\rout{r_{{\rm out}}}

\def\rvir{r_{{\rm vir}}}
\def\tvir{t_{{\rm vir}}}
\def\zvir{z_{{\rm vir}}}

\def\zrec{z_{{\rm recomb}}}

\def\omm{\Omega_{\rm m}}
\def\omb{\Omega_{\rm b}}
\def\lcdm{$\Lambda$CDM }

\def\Trei{T_{{\rm rei}}}
\def\Tprei{T_{{\rm prei}}}

\def\HI{{\rm H\,I}}
\def\HII{{\rm H\,II}}

\def\HeI{{\rm He\,I}}
\def\HeII{{\rm He\,II}}
\def\HeIII{{\rm He\,III}}

\def\LiI{{\rm Li\,I}}
\def\LiII{{\rm Li\,II}}
\def\LiIII{{\rm Li\,III}}
\def\LiIV{{\rm Li\,IV}}

\def\DI{{\rm D\,I}}
\def\DII{{\rm D\,II}}

\def\rhobg{\rho_{{\rm bg}}}

\title[Helium diffusion during structure formation]{Helium diffusion during formation of the first galaxies} 
\author[P. Medvedev et al.]{P. Medvedev$^{1}$\thanks{E-mail:
tomedvedev@iki.rssi.ru}, S. Sazonov$^{1,2}$,  M.~Gilfanov$^{3,1}$\\ 
$^{1}$Space Research Institute, Russian Academy of Sciences,
  Profsoyuznaya 84/32, 117997 Moscow, Russia\\ 
$^{2}$Moscow Institute of Physics and Technology, 9 Institutsky
  per., 141700 Dolgoprudny, Moscow Region, Russia\\
$^{3}$Max-Planck-Institut für Astrophysik, Karl-Schwarzschild-Str. 1,
  D-85741 Garching bei München, Germany}
\begin{document}

\date{2015}

\pagerange{\pageref{firstpage}--\pageref{lastpage}} \pubyear{2015}

\maketitle

\label{firstpage}

\begin{abstract}
We investigate the possible impact of diffusion on the abundance of helium and other primordial elements during formation of the first structures in the early Universe. We consider the primary collapse of a perturbation and subsequent accretion of matter onto the virialized halo, restricting our consideration to halos with masses considerably above the Jeans limit. We find that diffusion in the cold and nearly neutral primordial gas at the end of the Dark Ages could raise the abundance of primordial elements relative to hydrogen in the first virialized halos: helium enrichment could reach  $\delta Y_p/Y_p\sim 10^{-4}$ in the first star-forming minihalos of $\sim 10^5$--$10^6 \msun$. A moderate (to $\sim 100$~K) preheating of the primordial gas at the beginning of cosmic reionization could increase this effect to $\delta Y_p/Y_p\sim 3\times 10^{-4}$ for $\sim 10^6 \msun$ halos. Even stronger abundance enhancements, $\delta Y_p/Y_p \sim {\rm a~few~} 10^{-3}$, may arise at much later, post-reionization epochs, $z\sim 2$, in protogroups of galaxies ($\sim 10^{13}\msun$) as a result of accretion of warm-hot intergalactic medium with $T\sim 10^6$~K. The diffusion-induced abundance changes discussed here are small but comparable to the already achieved $\sim 0.1$\% precision of cosmological predictions of the primordial He abundance. If direct helium abundance measurements (in particular, in low-metallicity HII regions in dwarf galaxies) achieve the same level of precision in the future, their comparison with the BBN predictions may require consideration of the effects discussed here. 
\end{abstract}

\begin{keywords}
Primordial abundance, diffusion, early Universe, epoch of reionization
\end{keywords}

\section{Introduction}
\label{sec:Int}

In the very early Universe, 1 to 300 seconds after the Big Bang, neutrons and protons coupled into nuclei and the primordial plasma composition was formed: hydrogen, helium-4, deuterium, lithium-7 and a small admixture of other nuclides. Recently, measurements of the abundances of the primordial elements have drawn considerable attention. On the one hand, the predictions of the standard Big Bang nucleosynthesis theory (BBN) are becoming more precise. The only free parameter of BBN, the baryon-to-photon ratio, has been tightly constrained by observations of the cosmic microwave background (CMB) with the Planck space observatory. The Planck results and other recent cosmological data, together with  an improved nuclear reaction network, now constrain the primordial helium abundance to within a tenth of a per cent: $Y^{\rm BBN}_{\rm p} = 0.2463 \pm 0.0003$ \citep{coc14}. The corresponding predictions for the primordial abundances of deuterium and lithium are ${\rm D/H} = (2.65 \pm 0.07) \times 10^{-5}$ and ${\rm Li/H} = (4.9\pm  0.4) \times 10^{-10}$.

On the other hand, direct measurements of the primordial helium abundance have now reached a precision of about 1\%. The standard approach is based on observations of low-metallicity HII regions in dwarf galaxies and extrapolating the measured dependence of the helium abundance on metallicity to extremely low metallicities. Using this method, \citet{skillman13} achieved a precision of a few per cent: $Y^{\rm obs}_{\rm p} = 0.253 \pm 0.008$. Another recent determination by \citet{izotov13} provided an even more precise measurement: $Y^{\rm obs}_{\rm p} = 0.254 \pm 0.003$. 

One may hope that further progress in observational techniques will make it possible to improve the accuracy of direct helium abundance measurements by another order of magnitude or so and thus bring it to the already achieved $\sim 0.1$\% level of cosmological predictions of the primordial helium abundance, enabling a stringent test of the BBN paradigm. When dealing with such tight constraints it is crucial to account for any astrophysical phenomena that may have affected the helium abundance during the cosmic history. One such potentially important effect is considered in the present work. 

It has been demonstrated \citep{gilfanov84,shtykovskiy10,medvedev14} that long-term (billions of years) diffusion in the intracluster medium (ICM) of clusters of galaxies may cause a noticeable increase of the cluster-averaged helium abundance due to net inflow of helium from the intergalactic medium. Although this result was obtained for a hydrostatic model of the ICM immersed in an infinite reservoir of gas (assuming constant density at the outer boundary), a similar phenomenon  may be expected to take place in protoclusters and more generally during structure formation in the Universe. So far, there have been very few works on primordial element diffusion. \citet{medvigy01} studied diffusion of primordial elements in the linear regime of perturbation growth. \citet{kusakabe15} suggested that  ambipolar diffusion of ionized lithium during structure formation might explain the discrepancy between the observed lithium abundance (the so-called Spite plateau, \citealt{spite82}) and the BBN predictions, provided there are sufficiently strong primordial magnetic fields.

In this work we examine diffusion-driven separation of the primordial elements during the formation of the first star-forming minihalos and galaxies in the early Universe. We consider two stages of structure formation and evolution: i) the primary collapse of a halo and ii) subsequent accretion of matter onto the virialized halo. 

\section{Model}
\label{sec:model}
In astrophysical setups, diffusion can be caused by concentration, density and temperature gradients in a gas and by gravity. In many astrophysical problems, diffusion is considered for systems that are in a state of hydrostatic equilibrium. Typically, it means that the gravitational force is balanced by the opposite force of the gas pressure gradient, which is itself caused by the gravity. Therefore, if thermal diffusion is negligible, it is gravity that drives diffusion in such systems, giving rise to gravitational separation of different ionic species.

In our primary focus here is diffusion during the formation of the first structures in the Universe. In this case, an assumption of hydrostatic equilibrium would be appropriate only for rare objects with masses near the Jeans mass and also at the initial linear stage of growth of a perturbation of any mass. We do not consider this special case here and address the more common case of structures with masses considerably above the Jeans mass so that gas pressure is small compared to gravitational attraction. Essentially, we consider free fall of gas into a gravitational well. This, however, does not imply that there is no diffision. In fact, since there are still density and temperature gradients in the gas flows under consideration, atomic species with different masses and hence different thermal velocities experience different (albeit small) pressure forces per unit mass, which causes element diffusion. This type of diffusion is often referred to as barodiffusion (for a detailed discussion, see \S 58 in \citealt{landau87}). In contrast to the case of hydrostatic equilibrium, the diffusion here proceeds under ``zero gravity'', which means that the gravitational force drops from the equations governing diffusion velocities. In this case, the diffusion is slow compared to the bulk motion of the gas. 

We consider two stages of structure formation and evolution. The first stage is the primary collapse of a density perturbation including its linear and nonlinear phases, while the second one is the subsequent accretion of matter onto the virialized object. We first focus our attention on the early epoch between cosmological recombination and reionization when most of the baryonic matter in the Universe was neutral (Section~\ref{sec:cold_dif}). Next, we study diffusion in moderately heated, weakly ionized gas during the early stages of cosmic reionization (Section~\ref{sec:pre-heating}). Finally, we discuss effects of diffusion in the hot intergalactic gas after reionization of the Universe was completed (Section~\ref{sec:reionization}). 

Here are our basic assumptions:
\begin{itemize}

\item For both stages, we use a model of an isolated, spherically symmetric perturbation. For the second (accretion) stage, we use the self-similar solution of \citet{bertschinger85}. We assume that there is no magnetic field in the medium outside the forming halo (hereafter the intergalactic medium, IGM), which might be a reasonable assumption for the early cosmic epochs considered here.
 
\item We adopt a \lcdm cosmology with parameters inferred from the Planck data: $H_0 = 67.3\ \mbox{km}\ \mbox{s}^{-1} \ \mbox{Mpc}^{-1}$, $\omm = 0.3175$ and $\omb=0.049$ \citep{planck}. The present-day CMB temperature is $T_{{\rm CMB,0}} = 2.73$~K \citep{fixsen09}. We neglect the contribution of the $\Lambda$-term in equations of fluid dynamics, since we are interested in early epochs at $z\gg 1$. 

\item We assume that the IGM is an ideal gas with $\gamma = 5/3$. The initial abundance of elements is taken to be primordial and spatially uniform: the mass fraction of He$^4$ is $Y_p = 0.2463$; the abundances of deuterium and lithium are $A_{{\rm D}} = 2.65 \times 10^{-5}$ and $A_{{\rm Li^7}} = 4.9 \times 10^{-10} $, respectively \citep{coc14}. The latter are defined as the ratio of the number of nuclei of a given element to that of hydrogen. The baryon gas is modeled as a composition of 6 species: neutral hydrogen, neutral He$^4$, protons, singly or doubly ionized He$^4$, electrons and some minor species. The latter is either deuterium or lithium in various ionization states. Importantly, diffusion of the main species, hydrogen and helium, is not affected by the presence of minor species because of the very small abundance of the latter.

\item The free electron fraction $x_e$ is calculated with the RECFAST code \citep{seager}. The mean baryon temperature $T_m$ is obtained in the standard way from the balance equation taking into account the coupling of free electrons to the background radiation and the adiabatic expansion \citep{peebles93}. We also assume that gravitational compression induces an adiabatic rise in the gas temperature, so that $T = T_m (\delta+1)^{\gamma - 1}$, where $\delta(r,t)$ is the density contrast. In Sections~\ref{sec:pre-heating}--\ref{sec:reionization}, we additionally take into account IGM heating and ionization during cosmic reionization by explicitly setting $x_e$ and $T_m$.

\end{itemize}

\subsection{Stage I. Perturbation growth up to collapse}
\label{sec:stI}

We consider a spherical perturbation of radius $\ri$ and uniform overdensity $\di \ll 1$, or equivalently of mass $M_i$, at initial moment $\ti$ (the top-hat model). The well-known implicit solution for the motion of shells via conformal time parameter $\theta$ \citep[for details see][]{peebles80} is 
\begin{equation}
\label{eq:shells}
r= \frac{r_i}{\Delta_i} \sin^2 \frac{\theta}{2} \, ,
\end{equation}
\begin{equation}
\label{eq:shells2}
t= \frac{3}{4} \ti \Delta_i^{-3/2} (\theta - \sin \theta) \, ,
\end{equation}
where $r_i$ is the initial radius of a given shell and $\Delta_i$ is the initial overdensity within this shell. This solution holds both inside and outside the top-hat perturbation, with $\Delta_i = \di$ for $r_i \le \ri$ and $\Delta_i= \di \frac{\ri^3}{r_i^3}$ for $r_i > \ri$. 
We denote by $R(t)$ the solution of equations~(\ref{eq:shells}), (\ref{eq:shells2}) for the boundary of the perturbation, i.e. $R(\ti)=\ri$.

\begin{figure}
\includegraphics[width=1\columnwidth]{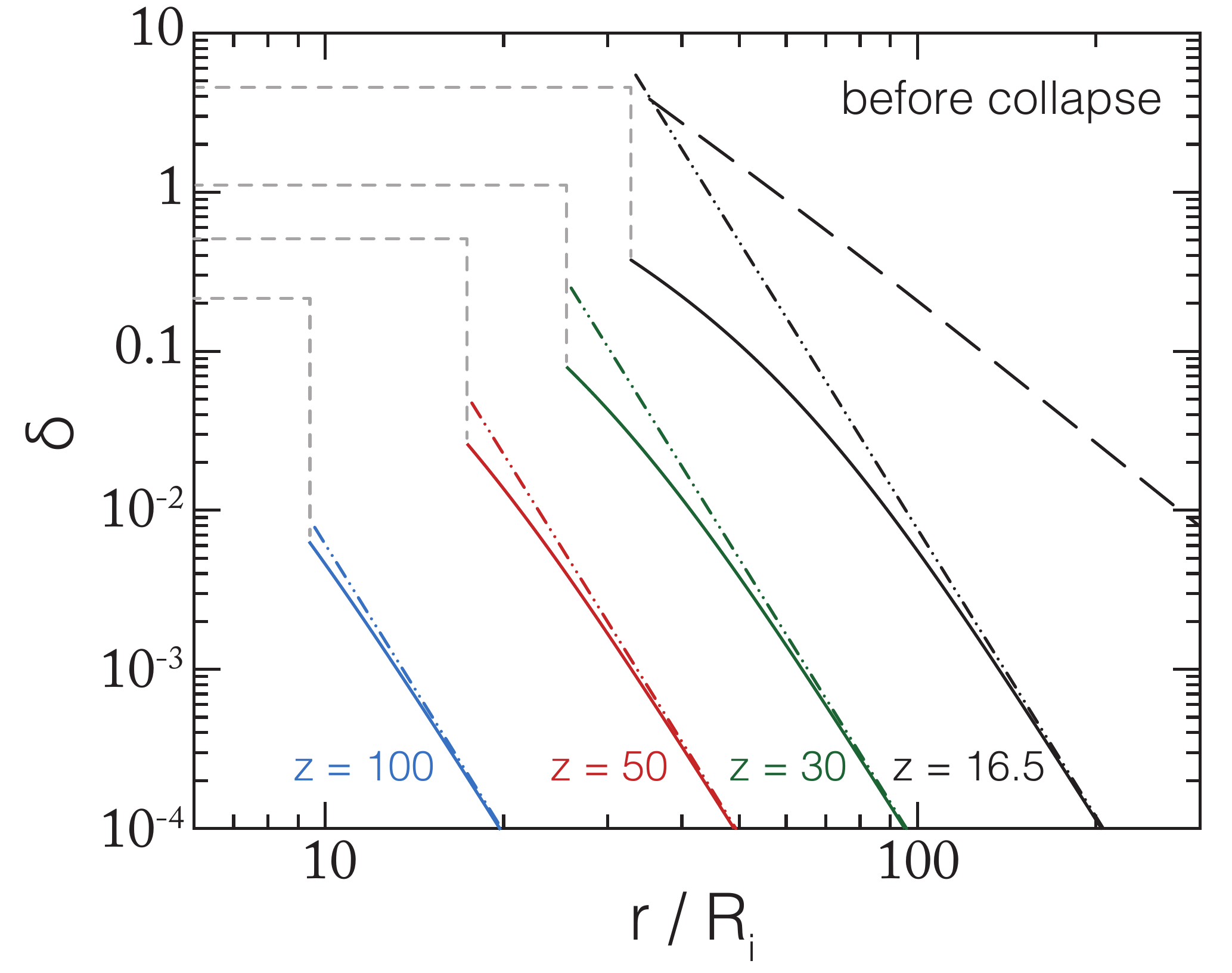}
\caption{Radial profiles of gas overdensity ($\delta = \rho/(\omb/\omm)\rhobg - 1$, eqs.~(\ref{eq:shells}), (\ref{eq:shells2}), (\ref{eq:mean_dens}), (\ref{eq:density_th})) inside (thin dashed lines) and outside (thick solid lines) of a growing spherical top-hat perturbation with $\mi = 10^8 \msun$ and virialization redshift $\zvir = 10$ at different moments: $z = 100$ (blue), $z = 50$ (red), $z = 30$ (green) and $z = \zita = 16.5$ (initial turn-around). The dash-dotted lines show an approximation valid in the outer parts of the density profiles (eq.~\ref{eq:apox_dens}). For $\zita$, also the radial profile of the mean density of gas interior to $r$ is shown by the thick long dashed line (eq.~\ref{eq:mean_dens}).}
\label{fig:stI}
\end{figure}

In this simplistic model shell crossing does not occur before collapse of the perturbation, and thus the mass interior to a given shell remains constant and is given by 
\begin{equation}
\label{eq:m_init}
M(t) = M_i = \frac{4}{3} \pi r_i^3 (1 + \Delta_i) (\rhobg)_i,
\end{equation}
where $(\rhobg)_i = \rhobg(\ti)$, $\rhobg = \rho_c \omm (1+z)^3$ is the background matter density and $\rho_c = 3 H_0^2 / (8\pi G)$ is the critical density. 
Assuming that baryons constitute a constant fraction of the total matter density, the mean gas density $\bar{\rho}$ within a given shell evolves as 
\begin{equation}
\label{eq:mean_dens}
\bar{\rho} = \frac{9}{2}\frac{\omb}{\omm}\rhobg \frac{(\theta - \sin \theta)^2}{(1 - \cos \theta)^3}.  
\end{equation}
By differentiating this equation, we can determine the radial profile of the gas density: for $r > R(t)$,
\begin{equation}
\label{eq:density_th}
\rho = \bar{\rho} \left[ 1 + 3 \left(1 - \frac{3}{2} \frac{\sin\theta (\theta - \sin\theta)}{(1 - \cos\theta)^2} \right)\right]^{-1};
\end{equation}
whereas for $r \le R(t)$, $\rho =\bar{\rho}$. To express the gas density as a function of physical quantities ($r$ and $t$), one ought to solve equations~(\ref{eq:shells}),  (\ref{eq:shells2}) for $\theta$ and substitute the result into equations~(\ref{eq:mean_dens}), (\ref{eq:density_th}). For $r \le R(t)$, the density is constant and $\theta$ can be obtained directly from equation~(\ref{eq:shells2}). For the flow outside the top-hat perturbation ($r > R(t)$), substitution of $r_i$ from equation~(\ref{eq:shells2}) into equation~(\ref{eq:shells}) yields:
\begin{equation}
\label{eq:r_th}
r =  \delta_i^{1/3} \ri \left(\frac{4}{3} \frac{t}{t_i} \right)^{8/9} \sin^2\frac{\theta}{2} (\theta - \sin \theta)^{-8/9}.
\end{equation}
By expanding equations~(\ref{eq:density_th}) and (\ref{eq:r_th}) at $\theta = 0$ and keeping the first nontrivial terms, we can approximate $\rho(r,t)$ as 
\begin{equation}
\label{eq:apox_dens}
\rho = (\omb/\omm) \rho_{bg} \left[ 1 + \frac{12}{175}  \di^{2}  \left(\frac{t}{\ti}\right)^{16/3} \left(\frac{r}{\ri}\right)^{-6} \right].
\end{equation}
Figure~\ref{fig:stI} shows how the gas density radial profile evolves during the growth of the perturbation.

The gas bulk flow velocity is
\begin{equation}
\label{eq:uI}
u(r,t) = \frac{dr}{dt} = \frac{r}{t} \frac{\sin \theta (\theta - \sin \theta)}{(1 - \cos \theta)^2}.
\end{equation}
It approaches the unperturbed Hubble flow velocity $v_H = 2r/3t$ as $\theta\to 0$. Assuming adiabatic gas compression, we can also determine the gas temperature profile:
\begin{equation}
\label{eq:T1}
T = T_m \left(\frac{\rho}{(\omb/\omm)\rhobg}\right)^{\gamma-1},
\end{equation}
where $T_m$ is the background gas temperature. Since the density increases with decreasing radius at $r>R(t)$, so does the temperature. 

The pressure gradient caused by the inward increase of density and temperature gives rise to barodiffusion in the gas flow outside the top-hat perturbation, while in the approximation used here there is no diffusion inside it (because the density and temperature are constant there). The perturbation ceases expanding and turns around to collapse at time $\tita = \frac{3\pi}{4} \di^{-3/2} \ti$ (corresponding to $\theta = \pi$), when its overdensity has increased to $\delta\approx 4.55$. We calculate net particle flows due to diffusion through the expanding outer boundary of the perturbation over the period from $\ti$ to $\tita$. Formally, the solution for spherical collapse remains valid up to a later time $\tvir = 2 \tita$, when the central density becomes infinite, but in reality perfect collapse is unlikely to happen due to the presence of anisotropies and angular momentum in the initial distribution of matter. Instead, the collisionless component (dark matter) is expected to reach virial equilibrium by ``violent relaxation'' \citep[for a detailed discussion see][]{mo10}, whereas the baryon gas will develop a shock and get heated to a temperature at which pressure balance will prevent further collapse. Due to the complexity of the relaxation process, we skip the virialization stage from our analysis. 

It should be noted that in the top-hat approximation used here, there is an abrupt jump of gas density at $R(t)$, seen in Fig.~\ref{fig:stI}. This formally implies an infinite diffusive flow through the boundary of the perturbation. However, this jump is outside our computational volume and does not affect the solution of the diffusion equations.

\subsection{Stage II. Cosmological accretion}
\label{sec:stII}

After virialization of the initial perturbation and formation of a bound halo, it continues to accrete matter. Shells outside the initially overdense region ($r_i>\ri$) turn around at successively later times, $\tita(r_i/\ri)^{9/2}$, so that the current turn-around radius at time $t$ is
\begin{equation}
\label{eq:rta}
\rta(t) = \rita \left(\frac{t}{\tita}\right)^{8/9},
\end{equation}
as can be found from equations~(\ref{eq:shells}), (\ref{eq:shells2}) by substituting $\theta = \pi$.

A 1-D model for halo formation by infall of matter in an expanding universe has been developed by \citet{gott75,gunn77,fillmore84,bertschinger85}. In the adiabatic solution, the baryon component accretes through an outwardly propagating shock near the virial radius, $r_{{\rm s}} \sim 1.3 r_{200}$.  Discussion of the cosmological infall is greatly simplified by the fact that the adiabatic solution is self-similar: all lengths can be scaled in terms of the current turn-around radius $\rta(t)$. In real situation, radiative gas cooling can lead to instabilities and disruption of the shock front at $r_{{\rm s}}$ and formation of a cooling flow \citep{forcado97,birnboim03}. Nevertheless, the adiabatic solution should still be valid at $r\gtrsim r_{{\rm s}}$. For simplicity, we refer to the radius of the shock in this solution as the virial radius, $\rvir \equiv r_{{\rm s}}$, and set this radius as the inner boundary of the free-fall region. Similarly to the first (perturbation growth) stage, we calculate net diffusive flows through this boundary, but this time in the post-virialization epoch $z \leq z_{{\rm vir}}$. Diffusion may also occur within the shocked gas at $r<\rvir$ and lead to redistribution of elements inside the halo (see, e.g., \citealt{gilfanov84,shtykovskiy10,medvedev14} for diffusion in the ICM), but consideration of this problem is beyond the scope of this work.

\begin{figure}
\includegraphics[width=1\columnwidth]{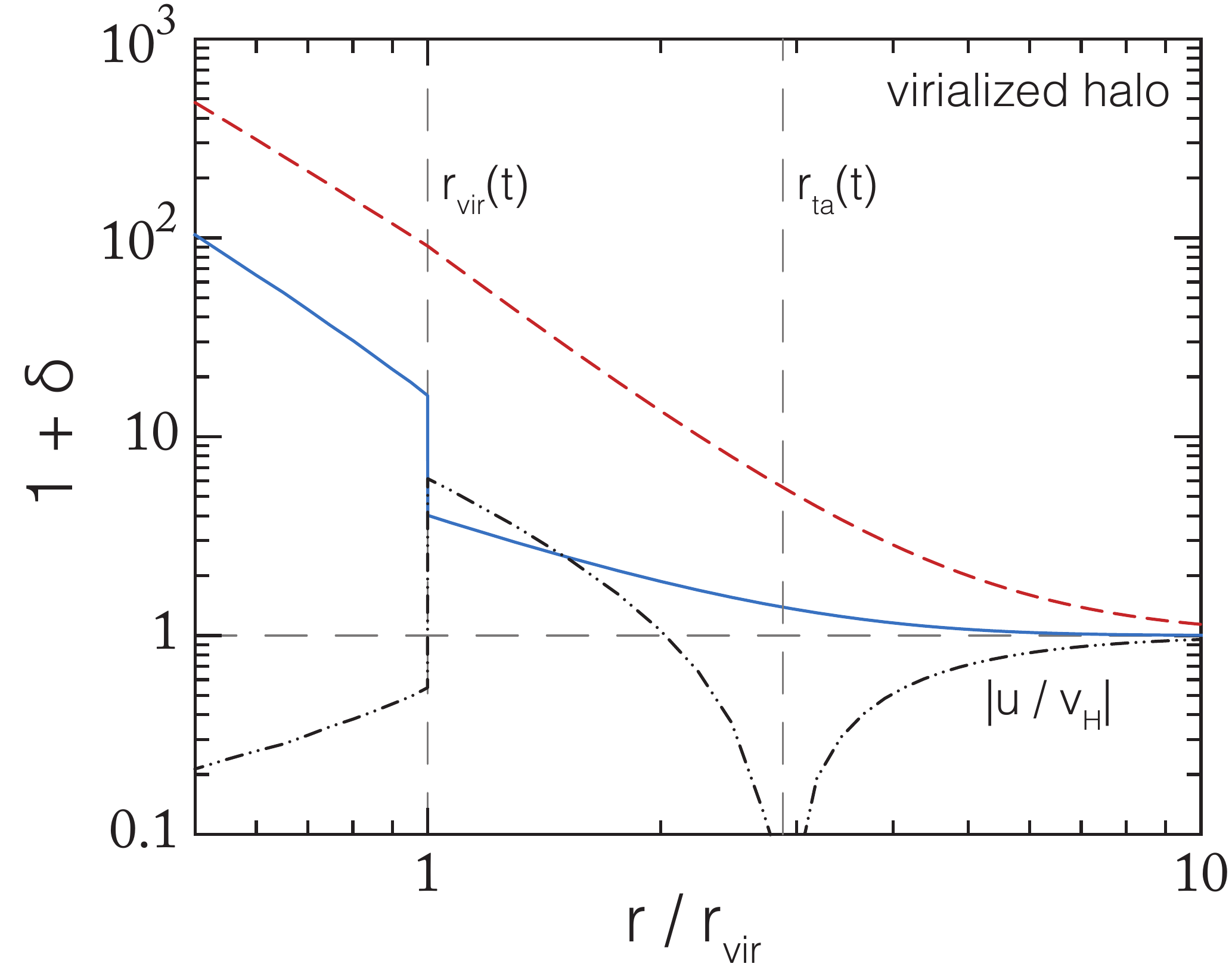}
\caption{Self-similar radial profiles of gas overdensity ($\delta = \rho/(\omb/\omm)\rhobg-1$, eq.~\ref{eq:density}, blue solid line), mean overdensity within $r$ (eq.~\ref{eq:mean_dens2}, red dashed line) and velocity (eq.~\ref{eq:u}, here divided by the Hubble velocity, black dash-dotted line) during accretion of matter onto a virialized halo at $z<\zvir$.}
\label{fig:stII}
\end{figure}

Following \citet{bertschinger85}, we can introduce the nondimensionless radius 
\begin{equation}
\label{eq:lambda}
\lambda \equiv \frac{r}{\rta(t)}
\end{equation}
and express the mean gas density within $r$ as 
\begin{equation}
\label{eq:mean_dens2}
\bar{\rho} = \frac{\omb}{\omm}\rhobg \lambda^{-3} \mathcal{M}(\lambda).
\end{equation}
The corresponding radial profiles of the gas density and velocity are
\begin{equation}
\rho = \frac{\omb}{\omm}\rhobg D(\lambda),
\label{eq:density}
\end{equation} 
\begin{equation}
\label{eq:u}
u = \frac{\rta}{t} V(\lambda).
\end{equation}
The functions $\mathcal{M}(\lambda)$, $D(\lambda)$ and $V(\lambda)$ for the dimentionless mass, density and velocity are tabulated in \citep{bertschinger85}. Assuming adiabatic gas compression, we can additionally determine the gas temperature profile:
\begin{equation}
\label{eq:T2}
T = T_m D^{\gamma-1}.
\end{equation}

For adiabatic gas with $\gamma = 5/3$, the shock occurs at fixed $\lambda \equiv \lambda_{{\rm vir}} = 0.347$ and propagates outwards as $\rvir\propto t^{8/9}$. Due to accretion, the total gas mass interior to $\rvir$ increases in time as $\propto t^{2/3}$. 
The pre-shock gas density and velocity are $D_{{\rm vir}} = 4.02$ and $V_{{\rm vir}} = -1.433$, which means that the pre-shock density is approximately four times the unperturbed density in the Hubble flow. The pre-shock mass parameter $\mathcal{M}_{{\rm vir}} = 3.799$ gives the mean overdensity within $\rvir$: $\bar{\rho} = 91(\omb/\omm)\rho_{bg}$.  As a consequence of self-similarity, the shape of the gas density profile is independent of time (see Fig.~\ref{fig:stII}).

\subsection{Calculation of diffusion}
\label{sec:calcdif}

We base our treatment of diffusion on the method developed by \citet{shtykovskiy10,medvedev14}. The main difference is that we now study a flow with a large Peclet number (the ratio of the advective transport rate to the diffusive one). Under this condition, diffusion does not affect macroscopic motions in the gas and can be regarded as an additive process to the bulk motion of the gas.

We use the same notation as in \citet[][section 3.1]{shtykovskiy10}. The gas is composed of N species with number densities $n_s$, masses $m_s$, charges $Z_s$ and mean velocities $u_s$, all having the same temperature $T_s = T$.  
The diffusion velocities $w_s$ are defined as:
\begin{equation}
w_s\equiv u_s-u, 
\label{eq:diffvelocity}
\end{equation}
where u is the mean fluid velocity (see eq.~(\ref{eq:u})):
\begin{equation} 
u=\frac{\sum_s n_sm_su_s}{\sum_s n_sm_s}.
\label{eq:meanfluidvel}
\end{equation}

Assuming spherical symmetry, the momentum transfer equation is
\begin{eqnarray}
&&n_s m_s \frac{D_s u_s}{Dt} + \frac{d(n_s k_B T)}{dr} + n_s m_s g - n_s Z_s e E =
\label{eq:moment}
\\
&&= \frac{\delta M_s}{\delta t}\equiv \sum\limits_{t\neq s} K_{st} [(w_t - w_s) + z_{st} (x_{st} r_s - y_{st} r_t)],
\nonumber
\end{eqnarray}
where $D_s/Dt$ is the convective derivative with respect to $u_s$, $g=G M(r)/r^2$ is the gravitational acceleration, $E$ is the radial electric field, $K_{st}$ is the friction coefficient between species $s$ and $t$ (see below), $r_{s,t}$ are the residual heat flow vectors \citep[for details see][]{burgers69,schunk75,thoul94} and $x_{st} = \mu_{st} / m_s,\ y_{st} = \mu_{st} / m_t$ with $\mu_{st} = m_s m_t / (m_s + m_t)$. We neglect the contribution of reactive collisions in the momentum equation (see for a discussion \citealt{geiss86}).

Since in the considered case $u \gg w_s$, the convective derivative can be approximated as $\frac{D_s u_s}{Dt} \approx \frac{D u}{Dt}$, where the latter derivative is defined with respect to the mean fluid velocity $u$ (see eq.~(17.15) in \citealt{burgers69}). This approximation essentially leads to elimination of gravity in equation~(\ref{eq:moment}), since the motion of the gas as a whole is governed by the Euler equation:
\begin{equation}
\frac{Du}{Dt} = -\frac{1}{\rho}\frac{dP}{dr} - g,
\end{equation}
where $P = k_BT\sum_s n_s$ is the total gas pressure.
Thus in this case the diffusion is driven by the small quantities ($\ll g$) remaining on the left-hand side of equation~(\ref{eq:moment}): 
\begin{equation}
\label{eq:burgers1}
\frac{dP}{dr} c_s \left( \frac{\eta m_p}{m_s} - 1\right) + P \frac{\eta m_p}{m_s} \frac{dc_s}{dr}  - n_s Z_s e E  = \frac{\delta M_s}{\delta t},
\end{equation}
where $\eta = 1/m_p \sum_s n_sm_s / \sum_s n_s$ is the mean molecular weight and $c_s = n_s m_s/\rho$ is the mass fraction of species $s$. Note that in  hydrostatic regime $\frac{D u}{Dt} = 0$ and then the largest quantity on the left-hand side of equation~(\ref{eq:moment}) is gravity, which is eliminated in our case.

We have the following equations for the residual heat flow vectors:
\begin{eqnarray}
&&\frac{5}{2} n_s k_B \frac{dT}{dr} = \sum\limits_{t\neq s}  K_{st} x_{st} \{ \frac{5}{2} z_{st} (w_s - w_t)
\label{eq:burgers2}
\\
&&- r_s y_{st} [ Y_{st}/x_{st}   +  \frac{4}{5} z_{st}^{\prime\prime}  ] + r_t  y_{st} [3 + z_{st}^{\prime}  - \frac{4}{5} z_{st}^{\prime\prime}]\}
\nonumber
\\
 &&- \frac{2}{5}  z_{ss}^{\prime\prime} r_{s} K_{ss},
\nonumber
\end{eqnarray}
where $Y_{st} = 3 y_{st} + z_{st}^{\prime} x_{st} m_t/m_s$. Here we also neglect terms related to the difference between the convective derivatives  $\frac{D_s u_s}{Dt}$ and $\frac{D u}{Dt}$ (see eq.~(17.24) in \citealt{burgers69}).

The friction coefficient between species $s$ and $t$ is
\begin{equation}
K_{st} = \frac{16}{3}  \mu_{st} \Omega_{st}^{(1,1)} n_s n_t,
\label{eq:K}
\end{equation}
where $\Omega_{st}^{(l,j)}$ is the Chapman-Enskog collision integral \citep{chapman70}. It can be related to the binary diffusion coefficient $D_{st}$ as
\begin{equation}
D_{st} = \frac{k_bT}{K_{st}} \frac{n_s n_t} { \sum_s n_s}.
\end{equation}
The coefficients $z_{st}$, $z_{st}^{\prime}$, $z_{st}^{\prime\prime}$ are the collision parameters, determined by $\Omega_{st}^{(1,1)},\Omega_{st}^{(1,2)},\Omega_{st}^{(2,2)}$ \citep[see][]{burgers69,schunk75}. 

Variation of the collision frequency with atomic species velocities causes so-called thermal diffusion (see, e.g., \citealt{monchick67}), which is essentially motion of more highly charged and more massive particles up the temperature gradient. For the problem at hand, due to our assumption of an adiabatic gas temperature profile, thermal diffusion operates in the same direction as barodiffusion, which leads to faster gravitational separation of hydrogen and heavier elements. However, thermal diffusion proves to be important only when the IGM is ionized and its temperature is higher than $10^4$~K. In other cases equation~(\ref{eq:moment}) can be simplified by omitting terms with $r_{s,t}$ and then equation~(\ref{eq:burgers2}) is no longer needed. In addition, there is no electic field for neutral gas, so $E=0$ in equation~(\ref{eq:moment}) in that case. 

\subsubsection{Neutral components}
\label{sec:crssecneutral}

We regard collisions between neutral atoms as collisions between rigid elastic spheres. In the range of energies of interest here, the neutral-neutral transport cross-section can be considered independent of energy. For interactions between hydrogen and helium atoms, we use the result of \citet{chung02}: $\sigma_{HeH} \approx 100$ a.u.. For interactions between neutral hydrogen and deuterium we use cross-sections from The Controlled Fusion Atomic Data Center\footnote{\texttt{http://www-cfadc.phy.ornl.gov/}} at energy 0.1 eV. For interecitons with lithium we adopt results from \citet{krupenie63}, which are the collision integrals at $T=1000$ K. The transport cross-sections for neutral-neutral collisions are listed in Table~\ref{t:neutral}. Since the cross-sections of identical particles, $\sigma_{ss}$, appear only in equation~(\ref{eq:burgers2}) for the heat flow vectors, these values are of little importance. As was mentioned before, thermal diffusion mainly operates in ionized hot gases.

The friction coefficient (or collision integral) between neutral particles is the transport cross-section $\sigma_{st}$ averaged over the microscopic velocity distribution:
\begin{equation}
K_{st} = \frac{16}{3} \left( \frac{k_B T \mu_{st} }{2 \pi}\right)^{1/2} \sigma_{st} n_t n_s.
\end{equation}
The corresponding collision parameters are $z_{st} = -1/5, z_{st}^{\prime} = 13/10, z_{st}^{\prime\prime} = 2$ \citep[see p. 182 in][]{burgers69}. 
\begin{table}
\topcaption{The transport cross-sections for neutral-neutral interactions in atomic units (1~a.u. $= 2.8 \cdot 10^{-17}\ \mbox{cm}^2$) adopted in this work}
\centering
\begin{tabular}{lcccc}\hline\hline
{$s$-$t$} & {H} & {He} & {D} & {Li}  \\
{H} & {194.1} & {100.0} & {88.6} &{223.5} \\  
{He}& {--} & {194.1} & {100.0} &{223.5} \\
{D}& {--} & {--} & {242.2} &{223.5}\\
{Li}& {--} & {--} & {--} &{381.0} \\
\hline\\
\end{tabular}
\label{t:neutral}
\end{table}

\subsubsection{Charged components}
\label{sec:crsseccharge}

In our energy range the cross-section for charged-neutral collisions tends to that given by the polarization approximation (``the Langevin cross section'',   see \citealt{pinto08} for comparison with full quantum mechanical calculations):
\begin{equation}
\sigma_{st} = 2.03 \times 10^{-15} Z_s^{1/2} \left(\frac{p_t}{\mbox{\AA}^3}\right)^{1/2}  \left(\frac{E_{st}}{\mbox{eV}}\right)^{-1/2}\ \mbox{cm}^2,
\end{equation}
where $p_t$ is the polarizability of neutral species $t$ and $E_{st}$ is the collision energy in the center-of-mass system. The polarizability of deuterium, hydrogen, helium and lithium are $p_D \approx p_{H} = 0.667 \mbox{\AA}^3, p_{He} = 0.207 \mbox{\AA}^3,p_{Li} = 24.31 \mbox{\AA}^3$ \citep{osterbrock61}.
The corresponding friction coefficient is given by
\begin{eqnarray}
K_{st} = 4.7 \times 10^{-33} Z_s^{1/2} (\mu_{st}/ m_p)^{1/2} \left(\frac{p_t}{\mbox{\AA}^3}\right)^{1/2} 
\nonumber
\\
n_s n_t\ \mbox{g~s}^{-1} \mbox{cm}^{-3}, 
\end{eqnarray}
where $m_p = 1.67 \times 10^{-24}$ g is the proton mass.
We treat the collision parameters for the interaction between charged and neutral particles as for the interaction between Maxwell-molecules \citep[see][]{schunk75}: $z_{st} = 0, z_{st}^{\prime} = 1, z_{st}^{\prime\prime} = 5/2$.

For the transport cross-section and the frictional coefficient of charged-charged collisions we use the standard expression of
the Coulomb momentum transfer rate \citep{spitzer56}:
\begin{eqnarray}
\label{eq:sigma_C}
\sigma_{st} &=& 2 \sqrt{\pi} e^4 Z_s^2 Z_t^2 (k_B T)^{-2} \ln{\Lambda_{st}},
\\
K_{st} &=& (2/3) \mu_{st} (2k_BT/\mu_{st})^{1/2} n_s n_t \sigma_{st}.
\end{eqnarray}
The Coulomb logarithm is assumed to be $\ln{\Lambda_{st}}= 22 + 3/2 \log (\frac{T_m}{30\ {\rm K}})$ everywhere \citep{burgers69}. We also assume that the collision parameters correspond to the pure Coulomb potential with a long-range cutoff at the Debye length \citep[see p. 182 in][]{burgers69}:  $z_{st} = 3/5, z_{st}^{\prime} = 13/10, z_{st}^{\prime\prime} = 2$.  

\subsection{Simulations}
\label{sec:simulations}

We first define the spatial region for our computations. Since we are interested in cumulative changes of element abundances for a given object, we are relatively free in the choice of the inner boundary, $\rin$. For stage I (pre-collapse), we define $\rin$ as the radius of the spherical perturbation $R(t)$, which keeps increasing until turn-around. For stage II (accretion onto the virialized halo), we set $\rin = \rvir \equiv r_{{\rm s}}(t) \propto t^{8/9}$. In both cases, we set a Neumann boundary condition at $\rin$, which means that we use a linearly extrapolated flux outside the computational domain to evaluate the derivative at the boundary. There is no shell crossing before collapse of the halo and thus no net gas flow across $\rin$, i.e. the mass of gas enclosed within $\rin$ remains constant during stage I. During stage II, the outer boundary of the halo expands while new gas accretes onto it, so that the halo's mass increases with time as $\propto t^{2/3}$. 

According to these definitions, the inner boundary of the diffusion region at the beginning of stage I, assumed to take place at $\zi = 10^3$, for a given initial halo mass $\mi$, is
\begin{equation}
\rin(\zi) \approx 8.5\ \mbox{pc}\ \left(\frac{\mi}{10^5 \msun}\right)^{1/3}.
\end{equation}
For stage II, for a given mass $M$ at $\zvir$, we find
\begin{equation}
\rin(\zvir) \approx 1.7\ \mbox{kpc}\ \left(\frac{M(\zvir)}{10^8 \msun}\right)^{1/3} \left(\frac{1 + \zvir}{11}\right)^{-1},
\end{equation}
which follows from equation~(\ref{eq:mean_dens2}). The redshift of virialization $\zvir \approx 0.6 \zita$ is a free parameter of the model (it is uniquely related to the initial overdensity $\di$ for stage I). 

We next define the outer boundary as $\rout \approx N \rin(\tf)$, where $\tf$ is the time when the calculation is stopped (for either stage I or stage II). We take this boundary distant enough that the IGM in its vicinity remains practically unperturbed and diffusion is negligible throughout the simulation. Hence, the outer boundary conditions are $w_{s}(\rout) = 0$, $u_s(\rout) = v_H$ (where $v_H$ is the Hubble velocity), $ \rho (\rout) = \omb/\omm \rho_{bg}$. Typically, $N \sim 5$ is sufficient, and if so, we verify that the solution is insensitive to $N$. 

We use a homogeneous Eulerian grid spanning from $\rin$ to $\rout$ with $N_p = 1000$ points. We perform computations using the standard forward-in-time, upstream (donor-cell) scheme. The number of grid points is chosen so as to ensure that numerical diffusion is negligible compared to physical diffusion. The time step is determined by the Courant criterion and is typically $dt \sim 10^{3}$--$10^{4}$~yr depending on the scale length of the problem, $dt \sim \frac{\rin}{N_p u(\rin)}$. Since the inner boundary of the grid changes with time, we need to adjust the grid at each time step: we reduce it by removing  cells whose central radius becomes smaller than $\rin(t)$. The outer boundary is fixed. 

Once the computational region is defined, we use the following scheme to calculate evolution of physical variables:
\begin{enumerate}

\item At given initial time $\ti$, the density $\rho$ (eq.~\ref{eq:density_th} or \ref{eq:density}), temperature $T$ (eq.~\ref{eq:T1} or \ref{eq:T2}) and mean fluid velocity $u$ (eq.~\ref{eq:uI} and~\ref{eq:u}) are derived for each gridpoint.

\item Using the initial abundances of elements and a given ionization fraction $x_e = n_e / n_{\rm H}$ (hereafter $n_{\rm H}$, $n_{\rm He}$, $n_{\rm D}$ and $n_{\rm Li}$ are the number densities of H, He$^4$, D and Li$^7$ nuclei, respectively; so that e.g. for hydrogen $n_{{\rm H}} = n_{\HI} + n_{\HII}$), we obtain the initial number densities of different species (ions or electrons), $n_s$. To this end, we use the primordial mass fractions of He$^4$, D and Li$^7$: $4 m_p n_{He} / \rho=0.2463$, $2 m_p n_{\rm D}/\rho=4.02\times 10^{-5}$ and $7 m_p n_{\rm Li}/\rho=2.6\times 10^{-9}$. The last two values have been derived from the abundances defined in a more usual way, $A_{\rm D}\equiv n_{\rm D}/n_{\rm H}$ and $A_{\rm Li^7}\equiv n_{\rm Li}/n_{\rm H}$, which were quoted in Section~\ref{sec:model}. Using mass fractions allows one to simplify the system of equations for the initial values of $n_s$ to a linear one. We also assume electro-neutrality for the gas: $ \sum_s Z_s n_s = 0$.

If the temperature is lower than $10^5$~K, we assume that the gas consists of $\HI$, $\HII$, $\HeI$, $\HeII$, electrons and some minor species. We assume that hydrogen and helium are ionized equally, i.e. $n_{\HII}/n_{{\rm H}} = n_{\HeII}/n_{{\rm He}}$. Obviously, for neutral gas $x_e = 0$ and there are no ionized species ($\HII$, $\HeII$, electrons). Similarly, for fully ionized gas there are no neutral particles ($\HI$, $\HeI$) and $x_e > 1$. When the temperature is higher than $10^5$~K, we assume that the gas consists of $\HII$, $\HeIII$, electrons and some minor species. Due to their very low abundance, we can perform computations for different minor species (${\rm D\,I}$--${\rm II}$, ${\rm Li\,I}$--${\rm IV}$) independently of each other. Adding a minor species to the gas mixture does not affect the diffusion of hydrogen or helium. 

\item We next solve Burgers' equations~(\ref{eq:burgers1}) and (\ref{eq:burgers2}) for the diffusion velocities $w_s$. To this end, we use two additional conditions needed to close the system: $\sum_s n_s m_s w_s = 0$ and $\sum_s Z_s n_s u_s = 0$. The former follows from the definition of $w_s$ and the latter from the absence of electric currents. 

\item Next, using $u$, $w_s$ and the boundary velocity $v_{{\rm in}} \equiv \partial \rin / \partial t$, we calculate particle flow rates through the inner boundary $\rin$: $f_s = \oint n_s v_s\,d{\bf i} = 4 \pi \rin^2 v_s n_s$, where $v_s = u + w_s - v_{{\rm in}}$. Note that for stage I, $u = v_{{\rm in}}$ and therefore $v_s = w_s$ at $\rin$, i.e. there is only diffusion and no bulk flow through $\rin$. The net amount of a given species accumulated in the halo over its evolution is obtained by time-integration of $f_s$.

\item The number density of a given species in the next time interval is obtained from the continuity equation: 
\begin{equation}
\frac{\partial{n_s}}{\partial{t}}+\frac{1}{r^2}\frac{\partial}{\partial{r}}[r^2n_s\,(w_s+u)]=0.
\end{equation}

\end{enumerate}

\section{Results}
\label{sec:results}

Using the method described above, we performed numerical calculations to estimate cumulative changes in the abundance of helium, deuterium and lithium that may arise due to diffusion during structure formation, as a function of redshift and halo mass.

\subsection{Minihalos fed by cold gas before reionization}
\label{sec:cold_dif}

We begin by considering the formation of the first bound objects in the Universe before it was reionized and reheated. Our model has two parameters: mass $\mi$ of the initial perturbation and the redshift of halo virialization $\zvir \approx 0.6\zita$. Since the ionized gas fraction after cosmic recombination is very small, $x_e \sim 10^{-4}$, the resulting diffusion velocities of hydrogen and helium prove to be nearly the same as they would be in the case of a completely neutral H-He gas. Due to the low temperature of the IGM and therefore small temperature gradients, thermal diffusion is not important in this case. Apart from thermal gas pressure, bulk motion of the baryons relative to the dark matter also counteracts accretion of gas into forming halos. This leads to the existence of a minimal halo mass for which baryons are still able to accrete onto halos. \citet{tseliakhovich11} have demonstrated that this filtering mass, averaged over the streaming velocity distribution, is of order $10^5 \msun$. We therefore adopt $10^5 \msun$ as the minimal halo mass in our computations, although the Jeans mass can be an order of magnitude smaller (see eq.~\ref{eq:MJ_cold} below).

\begin{figure*}
\includegraphics[width=160mm]{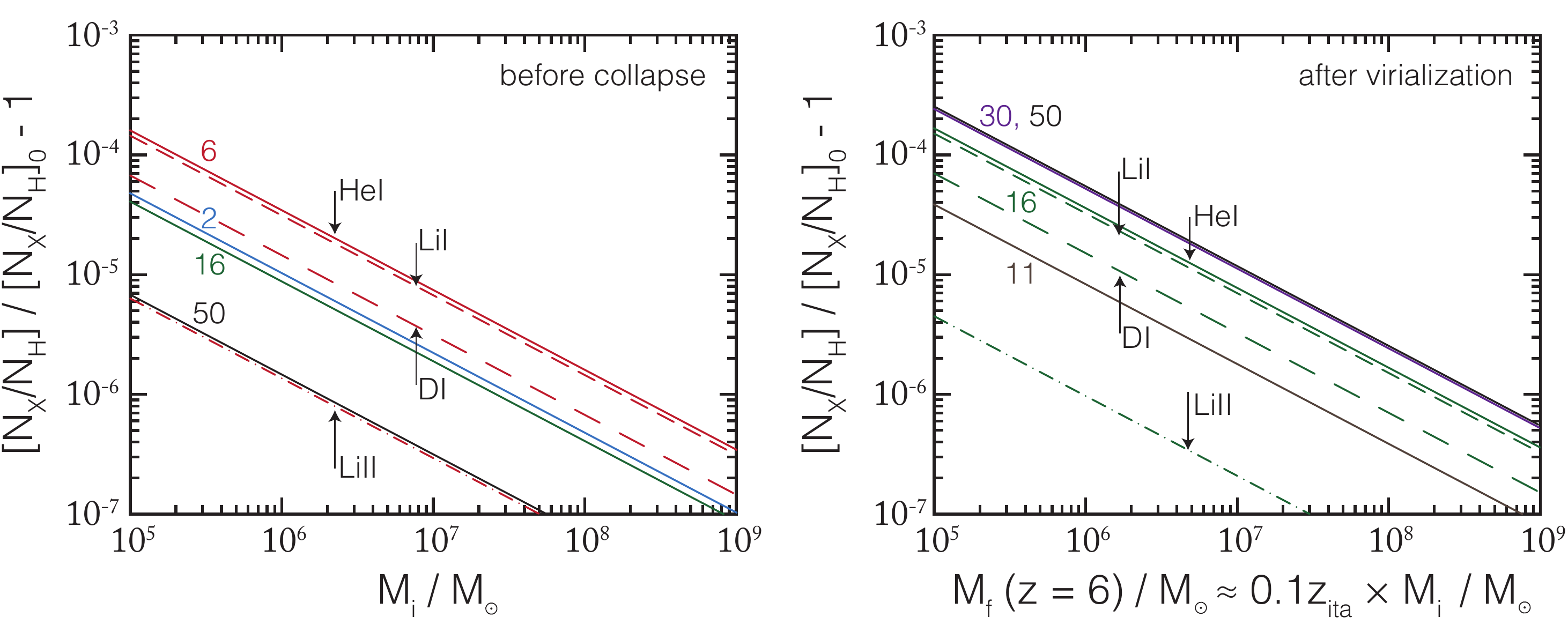}
\caption{Relative enhancements (per halo) of primordial element abundances caused by diffusion in the cold IGM before cosmic reionization. The solid lines correspond to $\HeI$, short dashed to $\LiI$, long dashed to $\DI$ and dash-dotted to $\LiII$. {\it Left}: Abundance increments accumulated during the growth of a spherical perturbation until its collapse (stage I) as a function of initial halo mass. The calculation starts at $\zi = 10^{3}$ and stops at either $\zita$ or the reionization redshift $\zrei = 6$, depending on which happens first. For $\HeI$, results for $\zita=2$ (blue), 6 (red), 30 (green) and 50 (black) are shown. For $\LiI$, $\DI$ and $\LiII$, only results for $\zita = 6$ are shown. {\it Right}: Abundance increments accumulated during secondary accretion onto the virialized halo (stage II) as a function of its final mass. Calculation starts at the virialization redshift $\zvir \approx 0.6 \zita$ and ends at $\zrei = 6$. For $\HeI$, results for $\zvir=6.5$ (brown), 10 (green), 18.5 (purple) and 31 (black) (corresponding to $\zita \approx$ 11, 16, 30 and 50) are shown. For $\LiI$, $\DI$ and $\LiII$, only the case of $\zvir=10$ ($\zita = 16$) is presented.}
\label{fig:cold_dif}
\end{figure*}

The left panel of Fig.~\ref{fig:cold_dif} shows the relative enhancement in the abundance of helium accumulated over the pre-collapse growth of a  perturbation (stage I in our model) as a function of $\mi$. We start our calculations at $\zi = 10^3$, i.e. just after cosmological recombintaion. The integration is stopped at $\zf$, the redshift of the earlier of two events: $\zita$, when the perturbation turns around to collapse, or the end of cosmic reionization $\zrei = 6$, when the Universe becomes completely ionized and our approximation of cold and nearly neutral IGM becomes invalid. 

We see that diffusion-driven helium enrichment diminishes with increasing halo mass and reaches a maximum amplitude $\delta Y_p/Y_p\sim 1.5 \times 10^{-4}$ for $\mi = 10^5\msun$ and $\zita \approx \zrei$. The decreasing trend with mass can be explained as follows. The instantaneous flux of particles of a given sort (e.g. helium) through the boundary $R(t)$ of the growing perturbation $q_s\propto R(t)^2 w_s(t)$. Due to self-similarity of the solution for spherical collapse, the diffusion velocity $w_s \propto \rho^{-1}\nabla P/P \propto \delta(1+\delta)^{-2}\rhobg^{-1} R^{-1}$ (times some temperature dependence arising from the diffusion coefficient). Here, $\rhobg(t)$ is the current mean matter density in the Universe and $\delta(t)$ is the current overdensity within the perturbation, which does not depend on the mass of the overdense region if the initial density contrast $\di(\ti)$ is fixed. The total mass gas (mainly hydrogen) within $R(t)$ remains constant with time ($M(t)=\mi$) and is $\propto (1+\delta)\rhobg R^3$. Therefore, diffusion is expected to change the helium abundance by $\delta (n_{\rm He}/n_{\rm H}) \propto \int q_s\,dt/M\propto \int \delta/(1+\delta)^{-3}\rhobg^{-2}R^{-2}\,dt \propto \mi^{-2/3}$, since $R\propto \mi^{1/3}$. The smaller effect for $\zita=2$ compared to $\zita=6$, observed in Fig.~\ref{fig:cold_dif}, is due to the fact that in this case the calculation was halted at $\zrei = 6$ when the perturbation was still growing and the pressure gradient was still small, accordingly limiting the diffusion velocity. 

Also shown in the left panel of Fig.~\ref{fig:cold_dif} are the corresponding abundance changes for minor species, namely $\DI$, $\LiI$ and $\LiII$, for $\zita = 6$, when diffusion is most efficient. Diffusion gradually raises all of these abundances in the forming halo, and the enhancements are comparable to or smaller than that for helium. Due to the low ionization potential of $\LiI$ ($\sim 5.39$~eV), nearly half of primordial lithium is expected to be in the singly-ionized state after cosmic recombination, at $\zrec > z > \zrei$ \citep{galli98}. The cross-section of $\LiII$--$\HII$ Coulomb interactions is much larger than that for $\LiII$--$\HI$ collisions. Even at $x_e \sim 10^{-4}$, this leads to spatial coupling of $\LiII$ and $\HII$ resulting in a net diffusive outflow of ionized lithium from the forming halo. However, the yet faster outflow of $\HI$ atoms leads to a slow increase of $\LiII$ abundance in the overdense region with time.

The right panel of Fig.~\ref{fig:cold_dif} shows the relative enhancements of the abundance of helium, deuterium and lithium accumulated during subsequent accretion of matter onto the newly formed halo (stage II) as a function of final halo mass $\mf$ at $\zrei = 6$ (where $\mf\approx (\zvir+1)/(\zrei+1)\times \mi$ at $z\gg 1$ for the self-similar accretion solution\footnote{As was mentioned before, we do not consider the virialization stage $\zita<z<\zvir$ and possible changes in the halo mass associated with this period.}). In this case, we start our calculation at the moment of virialization of the initial perturbation, $\tvir = 2 \tita$, and stop it at $\zrei = 6$ (assuming that $\zvir>\zrei$). 

As was the case with stage I, we see that the diffusion effect decreases with increasing halo mass. In addition, it increases with $\zvir$, due to the longer time available for operation of diffusion. Helium enrichment reaches a maximum of $\sim 2.5 \times 10^{-4}$ for $\mf = 10^5\msun$ when $\zvir\gtrsim 15$. The results presented in Fig.~\ref{fig:cold_dif} indicate that, for a given object, diffusion-induced abundance enhancements are typically 2--3 times larger for the accretion stage than for the pre-collapse stage. We thus focus our attention on the accretion stage in the subsequent discussion.

We conclude that diffusion could not raise the primordial abundances of helium, deuterium and lithium by more than $\delta A_{X}/A_{X}\sim 10^{-4}$ in the first bound structures formed in the Universe before it was reionized and experienced any preheating. Moreover, this effect is restricted to the smallest  halos of mass $\sim 10^5$--$10^6 \msun$. 

\begin{figure*}
\includegraphics[width=160mm]{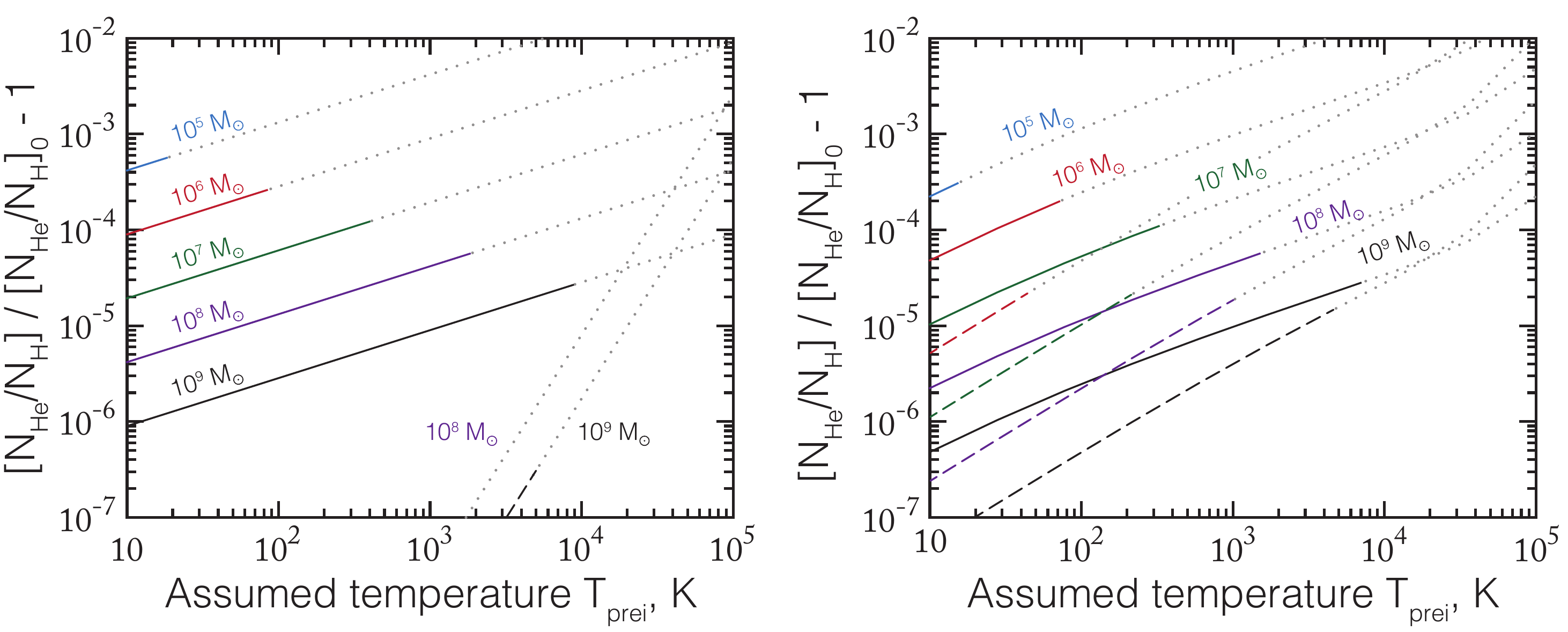}
\caption{Relative enhancement (per halo) of the helium abundance from its primordial value, caused by diffusion in the preheated IGM before reionizion, as a function of IGM temperature ($\Tprei$) for various ionization states. It is assumed that halos virialize at $\zvir=10$ and accrete matter until $\zrei = 6$, while the IGM temperature and ionization fraction remain constant. The stage of initial halo growth and collapse (before $\zvir$) is not taken into account. Shown are results for halos of $10^5$ (blue), $10^{6}$ (red), $10^{7}$ (green), $10^{8}$ (purple) and $10^{9}$ (black) $\msun$. The dotted parts of the curves indicate the unphysical regions where the mass of the object is smaller than the Jeans mass. {\it Left panel}: The solid lines correspond to neutral gas, while the dashed lines to fully ionized plasma. {\it Right panel}: The same, but for a different values of gas ionization: 20\% (solid), 90\% (dashed)}
\label{fig:Tas}
\end{figure*}

\subsection{Halos fed by preheated gas before reionization}
\label{sec:pre-heating}

Observations indicate that the IGM was almost fully reionized by $\zrei = 6$ \citep{spinrad98,hu99,fan00}. It is generally accepted that cosmic reionization was mostly driven by UV radiation from the first galaxies. However, it is likely (e.g., \citealt{venkatesan01,madau04,ricotti04}) that already during the early epoch of structure formation there existed X-ray sources that could significantly preheat and preionize the gas throughout the Universe. This might have been helped by cosmic-ray heating caused by the first supernovae \citep{sazonov15}. It is thus likely that the IGM was warm and weakly ionized long before $\zrei$, and such conditions could be favourable for diffusion. Therefore, we now examine the impact of IGM preheating on helium diffusion in halos forming before the reionization of the Universe was completed. 

We consider a simplistic scenario that at redshift $\zprei$ the IGM was suddenly preheated to a temperature $\Tprei$ and its ionization fraction was raised to $x_e$, and the IGM remained in this state until $\zrei$. As before, we study diffusion using the adiabatic solution for gas accretion onto a virialized halo, whose mass must exceed the Jeans mass \citep{shapiro94,barkana01}:
\begin{equation}
\label{eq:MJ}
\mj \equiv \frac{4\pi}{3} \left(\frac{\lambda_J}{2}\right)^3 \rhobg(0) \sim \frac{\pi}{6} \left( \frac{5\pi k_B T}{3 \eta m_p G}\right)^{3/2} \rhobg^{-1/2}.
\end{equation}
At $z \sim 150$, the mean baryon temperature decouples from that of the CMB and later on declines as $T_m \propto (1+z)^2$, and the corresponding Jeans mass 
\begin{equation}
\label{eq:MJ_cold}
\mj \approx 5 \times 10^3  \left(\frac{\eta}{1.23}\right)^{-3/2} \left(\frac{1+z}{10}\right)^{3/2}\ \msun ,
\end{equation}
where $\eta = 1.23\ (0.61)$ is the mean molecular weight for neutral (fully ionized) gas. If, however, the IGM is heated to a specified temperature $T_{{\rm IGM}}$, then 
\begin{equation}
\label{eq:MJ_hot}
\mj \approx 1.3 \times 10^{8} \left(\frac{\eta}{0.61}\right)^{-3/2} \left(\frac{T_{\rm IGM}}{10^3{\rm K}}\right)^{3/2}\left(\frac{1+z}{10}\right)^{-3/2}\ \msun,
\end{equation}

Figure~\ref{fig:Tas} shows the relative enhancement in the abundance of helium accumulated during accretion of a preheated IGM onto a virialized halo as a function of $\Tprei$. It is assumed that halos virialize at $\zvir=\zprei=10$ and accrete matter until $\zrei=6$. The stage of initial halo growth and collapse (before $\zvir$) is not taken into account. In the left panel, we illustrate the impact of heating on neutral and fully ionized gas. The large Coulomb cross-sections between charged particles effectively prevent diffusion in ionized plasma at low temperatures. However, the Spitzer diffusion coefficient steeply rises with temperature, much faster than for neutral gases: $D_{{\rm Coulomb}} \sim T^{5/2}$ vs. $D_{{\rm neutral}} \sim T^{1/2}$. As a result, for IGM temperatures above $10^4$ K the effects  of diffusion in fully ionized and neutral gases become similar. The right panel of Fig.~\ref{fig:Tas} shows results for partially ionized gas. If neutral particles are a dominant gas component, then neutral-charged interactions have small effect on diffusion of the neutral species, so that the effect of diffusion in weakly ionized gas ($x_e<40$\%) is similar to that in neutral gas.

We see that moderate IGM preheating can significantly increase the efficiency of diffusion of helium (and other primordial elements) during accretion of gas onto newly formed bound objects. For example, helium enrichment can reach $\delta Y_p/Y_p\sim 3\times 10^{-4}$ in minihalos of $\sim 10^6\msun$ for $\Tprei\sim 100$~K and $\delta Y_p/Y_p\sim 3\times 10^{-5}$ in more massive halos ($\sim 10^9\msun$) for $\Tprei\sim 10^4$~K, provided the ionization degree is not too high, $x_e\lesssim 40$\%. The former would correspond to objects that probably hosted the first, metal-free stars in the Universe, and the latter to the first galaxies in the Universe. 

\subsection{Massive halos fed by hot gas after reionization}
\label{sec:reionization}

We finally consider the case of virialized objects accreting hot ionized gas upon reionization of the Universe. We now assume that at $\zrei$ the IGM  became fully ionized ($x_e=1.08$--$1.16$, depending on whether helium is singly or doubly ionized) and acquired a constant high temperature $\Trei\gtrsim 10^4$~K. We further assume that there was a preceeding period $\zprei>z>\zrei$ of preheating when the IGM had lower temperature $\Tprei$ and ionization fraction $x_{e, \rm prei}$.

\begin{figure}
\includegraphics[width=1\columnwidth]{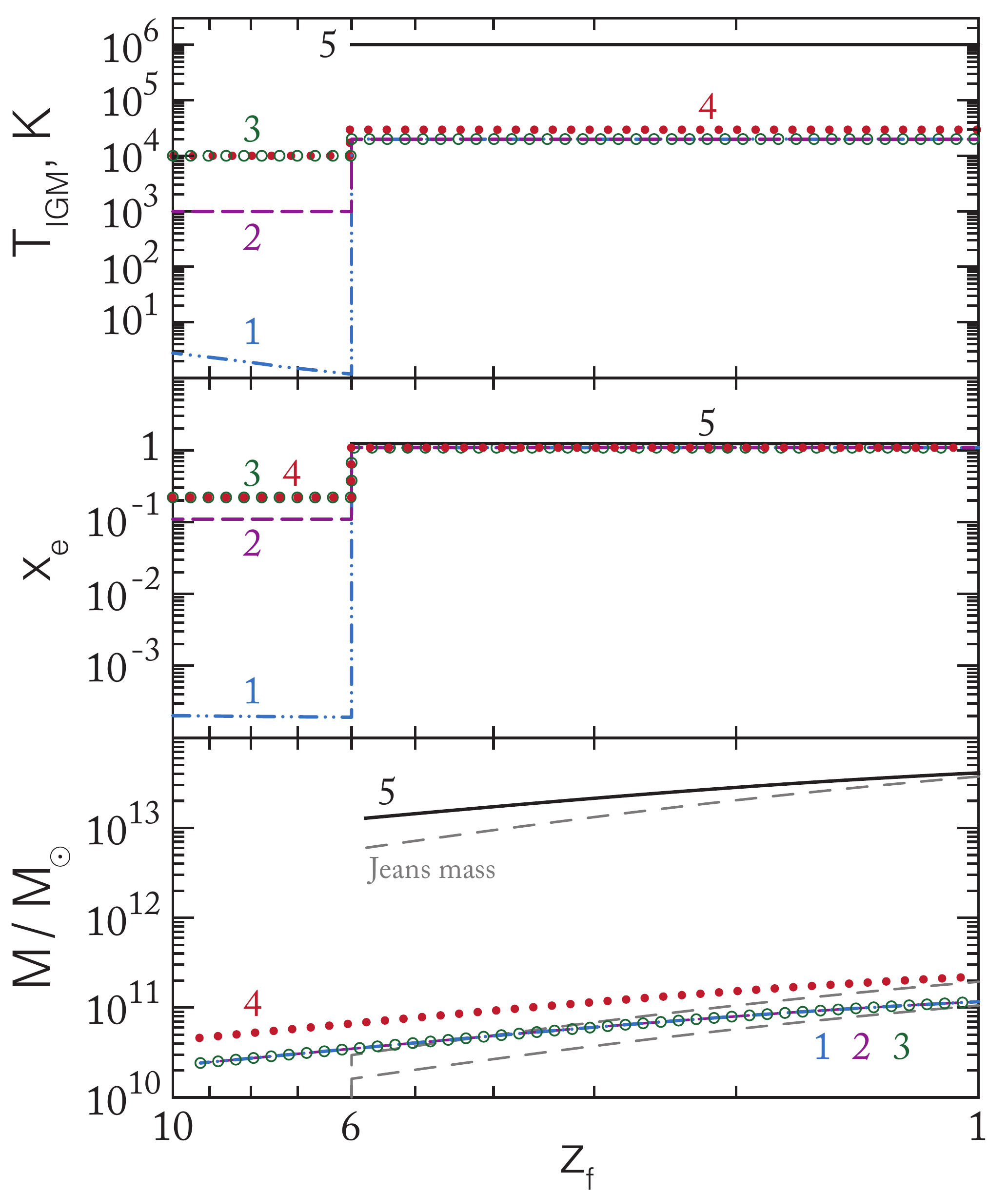}
\caption{Various scenarios of IGM reionization and heating considered here. The top and middle panels show the IGM temperature and ionization fraction as a function of redshift; the bottom panel shows the corresponding evolution of the Jeans mass (dashed gray lines) and of the mass of the halos used in diffusion calculations (the same types of lines as in the two upper panels). The computations started at $\zvir = 10$ for scenarios 1--4 and at $\zvir = 6$ for scenario 5. In all the scenarios, $\zrei = 6$; $\zprei = 10$ for scenarios 2--4 (see text for further details).}
\label{fig:scenarios}
\end{figure}

We calculated effects of diffusion in the post-reionization epoch for a number of such simplistic scenarios of IGM reionization:
\begin{enumerate} 
\renewcommand{\theenumi}{(\arabic{enumi})}
\item $\Trei = 2 \times 10^4$~K, without preheating; 
\item $\Trei = 2 \times 10^4$~K, $\Tprei = 10^3$~K and $x_{e, \rm prei} = 0.11$;
\item $\Trei = 2 \times 10^4$~K, $\Tprei = 10^4$~K and $x_{e, \rm prei} = 0.22$; 
\item $\Trei = 3 \times 10^4$~K, $\Tprei = 10^4$~K and $x_{e, \rm prei} = 0.22$;
\item $\Trei = 10^6$~K, no preceeding stage.
\end{enumerate}
We adopt $\zprei = 10$ and $\zrei = 6$. The scenarios are illustrated in Fig~\ref{fig:scenarios}.

\begin{figure}
\includegraphics[width=1\columnwidth]{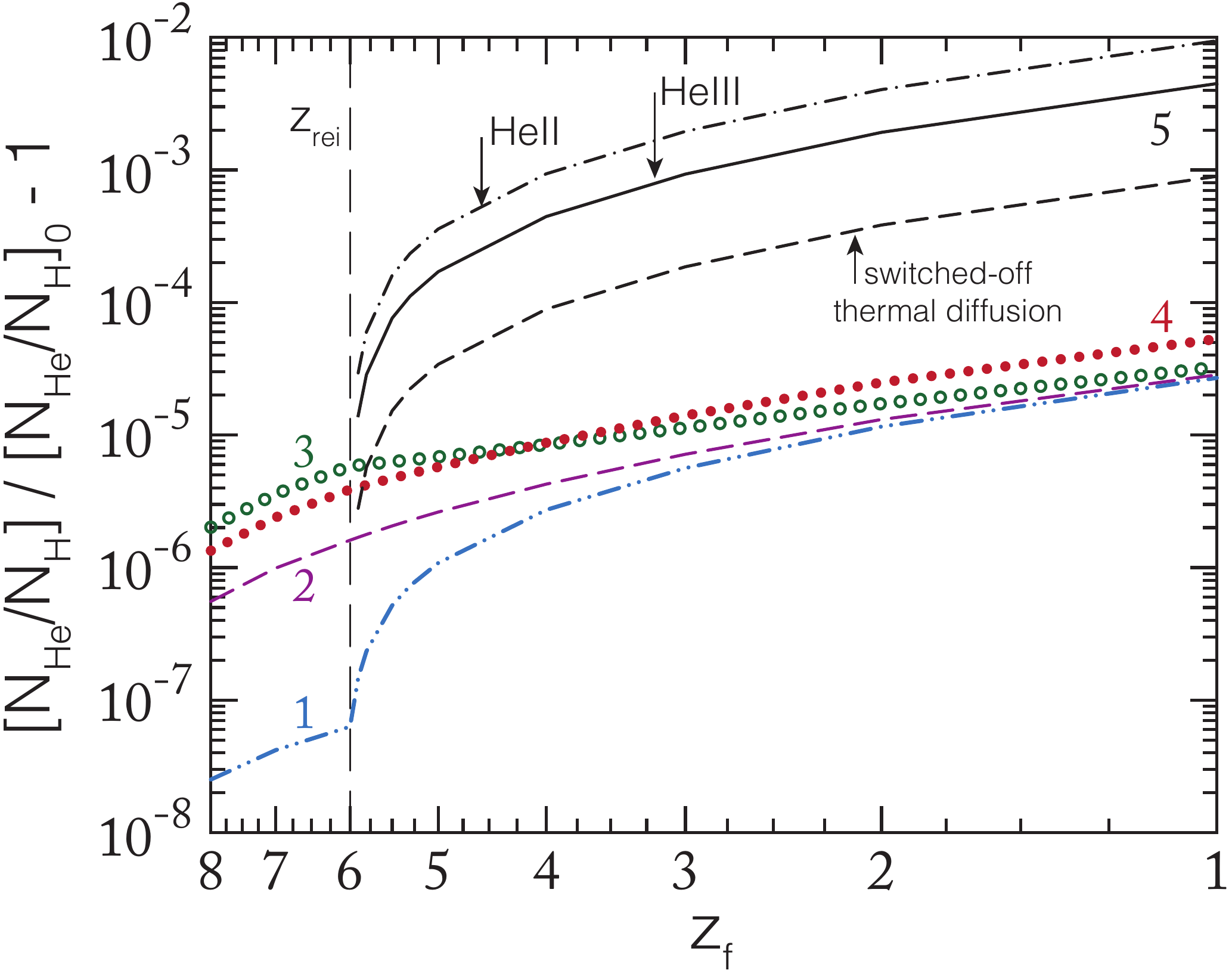}
\caption{Relative enhancement of helium abundance caused by diffusion, as a function of final redshift for scenarios described in Fig.~\ref{fig:scenarios}.  For scenarios 1--4, helium is assumed to be in the neutral and singly ionized states in the same proportion as hydrogen (determined by $x_e$). For scenario 5, we assume that all helium is $\HeIII$ (solid black), but show for comparison the result for $\HeII$ (black dash-dotted). Also shown is the result obtained for $\HeIII$ with thermal diffusion artificially turned off (grey dashed).}
\label{fig:rei}
\end{figure}

Figure~\ref{fig:rei} shows the results of these computations as a function of final redshift $\zf$. For scenarios 1--3, we start the calculations at the virialization redshift $\zvir = 10$ and consider halos with initial mass $M (\zvir)\approx 2\times 10^{10} \msun$ so that the halo remains above the Jeans limit during the post-reionization epoch. For scenario 4, we consider halos with $\zvir = 10$ and $M(\zvir) \approx 4\times 10^{10} \msun$. For scenario 5 (hot IGM), we take $\zvir = 6$ and $M(\zvir) \approx \times 10^{13} \msun$. In this case, we consider diffusion in a gas composed of $\HII$, electrons and either $\HeII$ or $\HeIII$.

We see that element diffusion during cosmological accretion of gas in the post-reionization epoch cannot change the abundance of helium in virialized halos by more than $\sim 10^{-5}$ if the accreting gas has a temperature of a few $10^4$~K, as expected for $z\lesssim 6$ \citep{benitez15}. However, a substantially stronger effect may be achieved in the case of accretion of hot plasma with $T\sim 10^6$~K, with thermal diffusion being important here (see Fig.~\ref{fig:rei}). In this case, helium enrichment may reach $\delta Y_p/Y_p \sim {\rm a~few~} 10^{-3}$ by $z\sim 2$. This pertains to objects of high (final) mass, $\sim 10^{13}\msun$, since smaller structures cannot accrete hot gas. While our scenario 5, where the IGM temperature is set at $10^6$~K already at $\zrei=6$, is certainly unrealistic, the change in the abundance of helium results from integration of the diffusion-driven inflow over the period $\zrei<z<\zf$ and is largely determined by what happens in its low-redshift part. Since a substantial fraction of the IGM at $z\sim 2$ is in fact expected to be in a warm-hot phase with $T\sim 10^5$--$10^7$~K \citep{cen99,cen06} and this gas should be accreting onto massive halos, a substantial enrichment of galaxies by helium during this epoch via diffusion might indeed be possible (provided magnetic fields do not change this simple picture completely). This deserves a more detailed assessment in future work. 

\begin{figure}
\includegraphics[width=1\columnwidth]{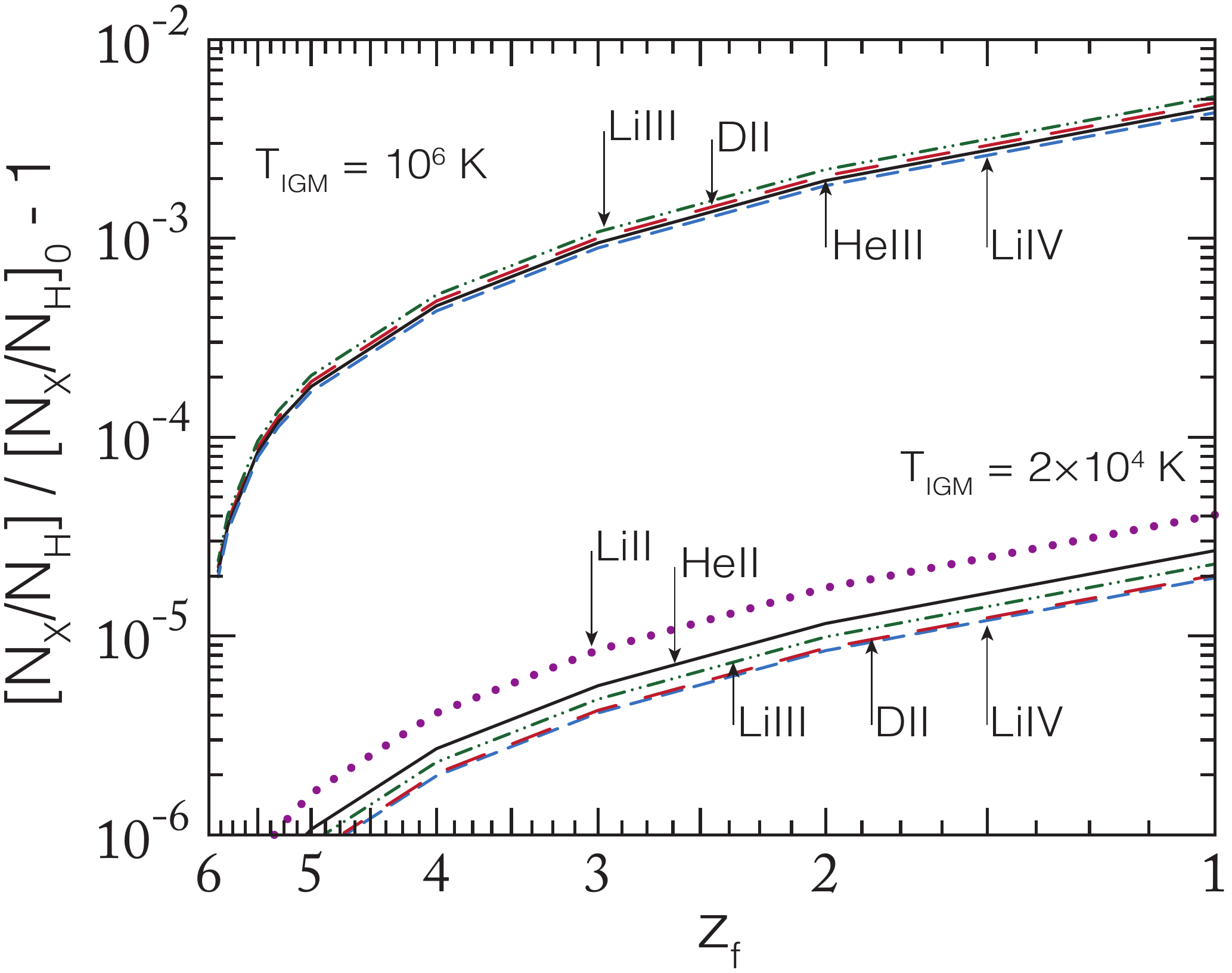}
\caption{Enhancement of deuterium and lithium abundances caused by diffusion during secondary accretion onto virialized halos during the post-reionization epoch, as a function of the final redshift. All calculations start at $\zvir \approx 6$. For the top group of curves, the IGM temperature $\Trei = 10^6$~K and the halo mass $M(\zvir) \approx \times 10^{13} \msun$. For the bottom group of curves, $\Trei = 2 \times 10^4$ K and $M(\zvir) \approx 2 \times 10^{10} \msun$. The short dashed lines are $\LiIV$, the long dashed ones are $\DII$, the dash-dotted ones  $\LiIII$, dotted curve is $\LiII$. For comparison, $\HeII$ and $\HeIII$ are shown by the solid lines.}
\label{fig:minor_sp}
\end{figure}

We can finally estimate the abundance changes for deuterium and lithium. Figure~\ref{fig:minor_sp} shows results obtained for two reionization temperatures $\Trei=2\times 10^4$~K and $10^6$~K. Here, for simplicity we started our calculations at $\zvir=\zrei=6$. We see that, as with helium, the largest abundance increments arise in the case of accretion of hot ($10^6$~K) plasma. The result also depends on the ionization state of a given element, primarily because barodiffusion depends on the mass and ion charge.   

\section{Summary}
\label{sec:summary}

We have explored the possible impact of diffusion on the abundance of helium and other primordial elements during formation of the first structures in the early Universe. We considered two stages of structure evolution: growth of a perturbation until its collapse and subsequent accretion of gas onto the virialized halo. 

At the end of the Dark Ages, when baryonic matter was cold and nearly neutral, diffusion could raise the concentration of He, D and Li  relative to hydrogen in the first virialized minihalos: the maximum enrichment for helium is $\delta Y_p/Y_p\sim 10^{-4}$ for $\sim 10^5$--$10^6 \msun$ halos. Moderate preheating of the IGM by X-ray irradiation or other mechanisms at the beginning of cosmic reionization could significantly accelerate diffusion during secondary accretion of gas onto virialized objects and accordingly change the chemical composition of their gas content. Helium enrichment could reach $\delta Y_p/Y_p\sim 3\times 10^{-4}$ in minihalos of $\sim 10^6\msun$, capable of forming the first stars, for an IGM heated to $\Tprei\sim 100$~K, and $\delta Y_p/Y_p\sim 3\times 10^{-5}$ in $\sim 10^9\msun$ halos, presumably hosting the first galaxies, for $\Tprei\sim 10^4$~K (provided the IGM remained moderately ionized, $x_e\lesssim 40$\%). Some of the protogalaxies formed at that time may still exist as dwarf galaxies at the present epoch and might be  the best targets to search for the predicted primordial abundance distortions.

It should be noted that the element abundances, including that of
helium, observed today in most places of the Universe have been
affected by stellar evolution. The effect of the latter is smallest in
low-metallicity dwarf galaxies, where helium abundance can be modified
by as little as $\delta Y_p/Y_p\sim 10^{-3}$ (whereas for galaxies with Solar
metallicity, $\delta Y_p/Y_p\gtrsim 10^{-2}$)
\citep{izotov13}. This is comparable, but yet somewhat larger than the
predicted amplitude of the effect of diffusion. Furthermore, for
measurements of primordial helium abundance, the effect of chemical
enrichment due to star formation is usually corrected for by
extrapolating the regression of the helium abundance versus
metallicity (typically, oxygen abundance) to zero metallicity
\citep{izotov13}.

We also examined accretion of gas onto virialized objects upon reionization of the Universe. In this case, diffusion cannot change the abundance of He, D and Li by more than $\sim 10^{-5}$ if the IGM has a temperature of a few $10^4$~K, as is expected for $z\lesssim 6$. However, a more noticebale helium enrichment, $\delta Y_p/Y_p \sim {\rm a~few~} 10^{-3}$, may be achieved by $z\sim 2$ in the case of accretion of warm-hot IGM with $T\sim 10^6$~K onto massive ($\sim 10^{13}\msun$) halos, corresponding to protogroups of galaxies. 

The diffusion-induced abundance changes discussed here are small but comparable to the already achieved $\sim 0.1$\% precision of cosmological predictions of the primordial He abundance. If direct helium abundance measurements (in particular, in low-metallicity HII regions in dwarf galaxies) achieve the same level of precision in the future (currently it is worse by an order of magnitude), their comparison with the BBN predictions may require consideration of the effects discussed here. 

\section*{ACKNOWLEDGMENTS}
PM and SS acknowledge the Russian Science Foundation for support of this work through grant 14-12-01315 and also thank the Max-Planck-Institut f\"ur Astrophysik for its hospitality.

\end{document}